\documentclass[prx,twocolumn,amsmath,amssymb,
reprint,%
author-year,%
author-numerical,%
superscriptaddress,
dvipdfmx
]{revtex4}

\usepackage{graphicx}
\usepackage{bm}
\usepackage{color}

\begin{document}


\title{Simple prediction of immiscible metal alloying based on metastability analysis}

\author{Shota Ono}
\email{shota\_o@gifu-u.ac.jp}
\affiliation{Department of Electrical, Electronic and Computer Engineering, Gifu University, Gifu 501-1193, Japan}

\author{Junji Yuhara}
\affiliation{Department of Energy Engineering, Nagoya University, Nagoya 464-8603, Japan}
\author{Jun Onoe}
\affiliation{Department of Energy Science and Engineering, Nagoya University, Furo-cho, Chikusa-ku, Nagoya, Aichi 464-8603, Japan}

\begin{abstract}
It has been known that even though two elemental metals, $X$ and $Y$, are immiscible, they can form alloys on surfaces of other metal $Z$. In order to understand such surface alloying of immiscible metals, we study the energetic stability of binary alloys, $XZ$ and $YZ$, in several structures with various coordination numbers (CNs). By analyzing the formation energy modified to enhance the subtle energy difference between metastable structures, we find that $XZ$ and $YZ$ with B2-type structure (CN$=$8) become energetically stable when the $X$ and $Y$ metals form an alloy on the $Z$ metal surface. This is consistent with the experimental results for Pb-Sn alloys on metal surfaces such as Rh(111) and Ru(0001). Some suitable metal substrates are also predicted to form Pb-Sn alloys. 
\end{abstract}

\maketitle

\section{Introduction}
Characterizing the structure of alloys is an important issue in materials science. In general, alloys can be classified into two groups: ordered alloys having regular lattices and disordered alloys (or solid solutions). On the other hand, some metals are immiscible with each other in the bulk. Therefore, many attempts have been made to understand the structure of alloys in a unified manner. For example, the use of the Mendeleev number has enabled us to categorize the structure of binary alloys \cite{pettifor}. More recently, in order to predict the ground-state structures, the density-functional theory (DFT) approach combined with machine learning methods \cite{ceder,schleder}, high-throughput calculations \cite{wolverton,hart}, and cluster expansion methods \cite{nelson_CE,seko} has been proved to be useful.  

The surface alloying has been observed when bulk alloys can be synthesized \cite{overbury,dhaka,sad}. Counterintuitively, the surface alloying has also been reported between immiscible elements \cite{nielsen,roder,nagl,steve,tober,sadigh,chen,yuhara_Rh,yuhara_Ru,yuhara_Ag}. One of the authors has investigated the alloying of immiscible metals (Pb and Sn) on various surfaces including Rh(111) \cite{yuhara_Rh}, Ru(0001) \cite{yuhara_Ru}, Ag(111) \cite{yuhara_Ag}, and Al(111) \cite{yuhara_Al}. On the Rh and Ru surfaces, the Pb-Sn films form ordered structures; on the Ag surface, the Pb-Sn films form disordered structures; and on the Al surface, the Pb and Sn atoms are immiscible. These indicate that the substrate plays a crucial role in determining the structure of Pb-Sn thin films. While the formation of surface alloys, in principle, can be studied by using DFT approach \cite{yang,marathe2009,marathe2013}, it would be a very complex task to treat surfaces with many atoms.  

The coordination number (CN) would be a key to understand the alloying of immiscible metals. Nielsen et al. have reported the growth of Au on Ni(110), irrespective to their immiscibility in the bulk \cite{nielsen}. Within the effective medium theory, they have shown that the cohesive energy of Au as a function of the number of Ni neighbors takes the minimum when Au atom is surrounded by eight Ni atoms. This CN is smaller than twelve realized in the face-centered cubic lattice structure, but is similar to the number, seven, realized in the Ni(110) surface, yielding an increase in the energy gain when alloyed near the surface. This study implies that the substrate effect can be incorporated in the CN. We expect that this is also modeled by the energetic stability of metastable (or hypothetical) structures: the total energies of alloys in various structures would provide a useful information for understanding the alloying. 

\begin{figure*}[tt]
\center
\includegraphics[scale=0.68]{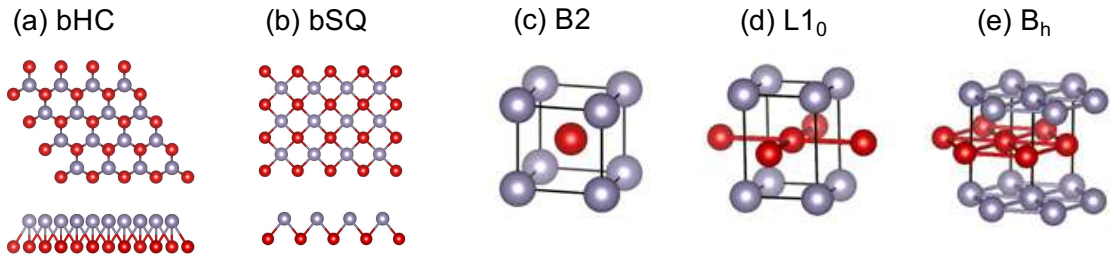}
\caption{The crystal structures of (a) bHC (top and side), (b) bSQ (top and side), (c) B2, (d) L1$_0$, and (e) B$_h$ alloys.} \label{fig_structure} 
\end{figure*}

In this paper, we explore a simple scheme for predicting the alloying of immiscible metals on another metal surface based on DFT calculations with less computational cost. In order to understand the formation of alloys $XY$ on the $Z$ metal substrate, we study the energetic stability of binary alloys $XZ$ and $YZ$ in five structures including buckled honeycomb (bHC), buckled square (bSQ), B2, L1$_0$, and B$_h$ (see Fig.~\ref{fig_structure}). These have different CN, allowing us to study the substrate effect. We propose the modified formation energy that identifies the effect of different CNs on the alloying, and demonstrate that if the alloy $XY$ on the surface $Z$ has been synthesized experimentally, the modified formation energy of the B2 structure (CN$=$8) is negatively large. Using this fact and our high-throughput calculations \cite{ono2020}, we predict suitable substrates for synthesizing Pb-Sn alloys. We also demonstrate that the metastability of strain-induced alloys behaves different manner, where the B2 does not take the minimum value of the modified formation energy.  

The advantage of the present approach is to predict suitable substrates for the surface alloying without performing the structure optimization of slab models. The physics behind this simplification is that the B2 structure model partly accounts for the interaction between the surface alloy and the substrate, in the sense that the CN in both systems is similar value. On the other hand, it is difficult to predict which surface orientations are suitable for alloying. In this way, we consider that the present approach can be used for screening the combination of elemental metals. It would be desirable to determine the complex structure of surface alloys predicted below and study its dynamical stability by performing phonon dispersion calculations. However, it is beyond the scope of the present work.  

\section{Computational details}
We calculated the total energy of alloys based on DFT implemented in \texttt{Quantum ESPRESSO} (\texttt{QE}) code \cite{qe}. The computational details were the same as in the high-throughput calculations \cite{ono2020}. We used the Perdew-Burke-Ernzerhof exchange-correlation functional within the generalized gradient approximation \cite{pbe} and used the ultrasoft pseudopotentials (\texttt{pslibrary.1.0.0}) \cite{dalcorso}. The cutoff energies for the wavefunction and the charge density were set to 80 and 800 Ry, respectively. The self-consistent calculations within spin-restricted approximation were performed by using 20$\times$20$\times$1 $k$ grid and 20$\times$20$\times$20 $k$ grid for two-dimensional (2D) and three-dimensional (3D) structures, respectively \cite{MK}. The smearing parameter of Marzari-Vanderbilt \cite{smearing} was set to $\sigma=0.02$ Ry. For 2D structures, we set the size of the unit cell along the $c$ axis to be 14 \AA \ that is enough to avoid the interlayer coupling between different unit cells. The total energy and forces were converged within $10^{-4}$ Ry and $10^{-3}$ a.u., respectively. 

The standard formation energy of the binary alloy $XY$ in the structure $j$ is defined as
\begin{eqnarray}
 F_j(XY) = \varepsilon_j(XY)- \frac{1}{2}\left[\min_j\varepsilon_j(X) + \min_j \varepsilon_j(Y)\right],
\label{eq:Eform1}
\end{eqnarray}
where the values of $\varepsilon_j(XY)$ is the total energy of alloy $XY$ in the structure $j$. $\min_j \varepsilon_j(X)$ is the minimum value of $\varepsilon_j(X)$ among $j$s, where $\varepsilon_j(X)=\varepsilon_j(XX)$. Negative value of $F_j(XY)$ indicates that alloying yields energetically more stable structure. However, with the use of Eq.~(\ref{eq:Eform1}), it would be difficult to distinguish subtle energy difference (a few meV per atom) among 3D (and 2D) structures. Alternatively, by adding and subtracting the total energies of the structure $j$ for elements $X$ and $Y$, Eq.~(\ref{eq:Eform1}) can be seen as
\begin{eqnarray}
 F_j(XY) = E_j(XY)+\frac{1}{2}\left[ S_j(X)+S_j(Y) \right],
\end{eqnarray}
where $E_j(XY)$ is the modified formation energy defined as 
\begin{eqnarray}
 E_j(XY) = \varepsilon_j(XY)- \frac{1}{2}\left[\varepsilon_j(X) + \varepsilon_j(Y)\right],
\label{eq:Eform2}
\end{eqnarray}
and $S_j(X)$ is the structure energy defined as
\begin{eqnarray}
 S_j(X) = \varepsilon_j(X) - \min_j\varepsilon_j(X).
\end{eqnarray}
When the element $X$ has the structure $j=G$ as its ground state, $S_G(X)$ is zero exactly, yielding $F_G(XY)=E_G(XY)$. For the metastable structure $M$, $S_M(X)>0$ by definition. Therefore, the positive values of $S_M(X)$ and $S_M(Y)$ must be cancelled out by negative value of $E_M(XY)$. This is because the value of $E_M(XY)$ is measured from the energy of metastable structure of $X$ and/or $Y$, as in Eq.~(\ref{eq:Eform2}). The energy difference between 3D (and 2D) structures in Eq.~(\ref{eq:Eform2}) becomes much larger than that in Eq.~(\ref{eq:Eform1}), which will be useful to identify the CN-dependence of the formation energy. For example, if the elements $X$ and $Z$ in the B2 (i.e., the body-centered cubic) structure are less stable but the alloy $XZ$ in the B2 structure is energetically stable, $E_{\rm B2}(XZ)$ will be negatively large, implying that the surface alloying with CN$=$8 is preferred. Below, we used Eq.~(\ref{eq:Eform2}) as the formation energy. 

The CNs of bHC, bSQ, B2, L1$_0$, and B$_h$ structures are three, four, eight, twelve, and twelve, respectively. The latter two values may depend on the ratio $c/a$ of the lattice parameters: when $c/a=\sqrt{2}$ and 1.63 in L1$_0$ and B$_h$, respectively, the CNs are twelve exactly. On the other hand, when $c/a=1$ exactly in L1$_0$, such a structure is the same as the B2 structure. 

\section{Results and discussion}

\begin{figure}[tt]
\center
\includegraphics[scale=0.5]{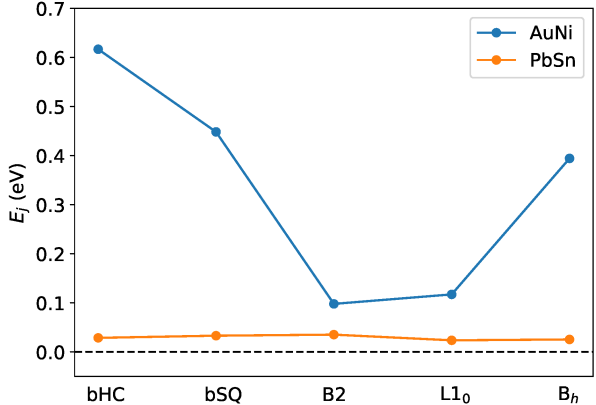}
\caption{The formation energy of AuNi and PbSn in the bHC, bSQ, B2, L1$_0$, and B$_h$ structures.  } \label{fig_PbSn} 
\end{figure}

\begin{figure}[tt]
\center
\includegraphics[scale=0.5]{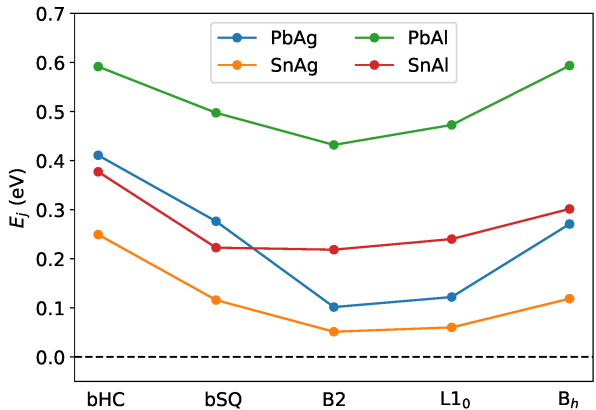}
\caption{Same as Fig.~\ref{fig_PbSn} but for PbAg, PbAl, SnAg, and SnAl. } \label{fig_Ag_Al} 
\end{figure}

\begin{figure}[tt]
\center
\includegraphics[scale=0.5]{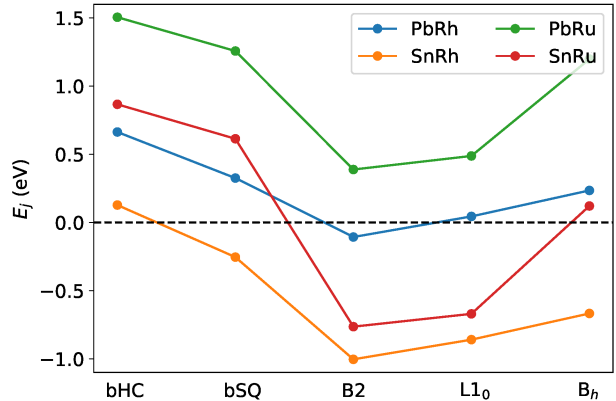}
\caption{Same as Fig.~\ref{fig_PbSn} but for PbRh, RbRu, SnRh, and SnRu.  } \label{fig_Rh_Ru} 
\end{figure}

As mentioned, Au is immiscible with Ni in the bulk but forms alloys on the Ni surface \cite{nielsen}. In order to demonstrate how our approach captures this alloying, we first consider the stability of Au-Ni systems. Figure \ref{fig_PbSn} shows the values of $E_j({\rm AuNi})$ for $j=$bHC, bSQ, B2, L1$_0$, and B$_h$ structures (blue circles). Among five structures, the B2 structure has the lowest $E_j$. Since CN$=$8 for B2, this is consistent with the effective medium theory analysis \cite{nielsen}, indicating that the surface alloying is related to the relative stability of B2 structure.

\begin{figure*}
\center
\includegraphics[scale=0.52]{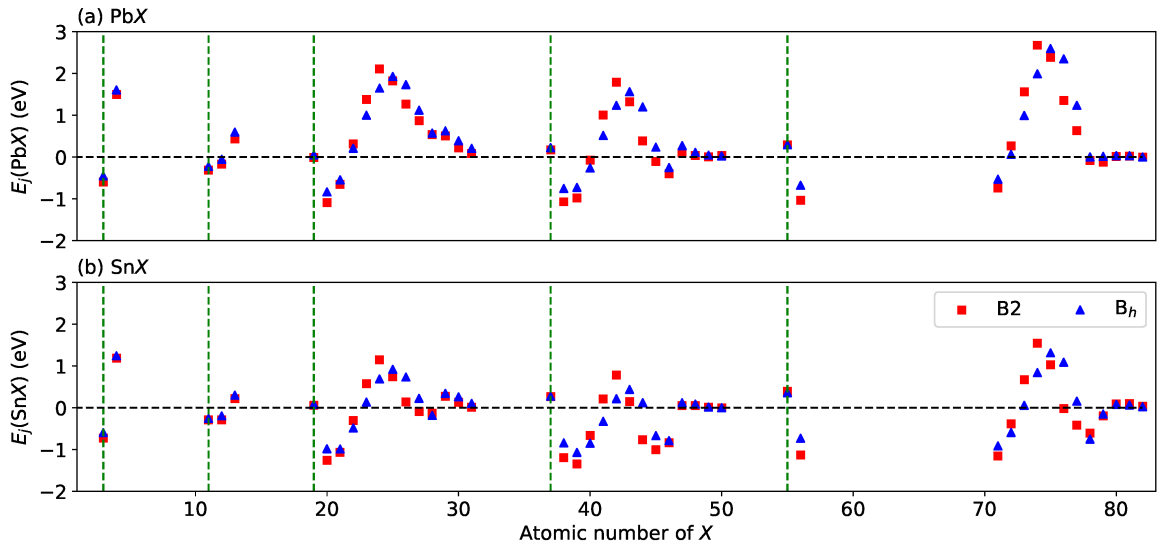}
\caption{$E_j$ as a function of the atomic number of $X$ for (a) Pb$X$ and (b) Sn$X$ alloys in the B2 (square) and B$_h$ (triangle) structures. The vertical dashed lines indicate the atomic number of alkali metals (Li, Na, K, Rb, and Cs). } \label{fig_X} 
\end{figure*}

\subsection{Pb-Sn alloys}
We next consider the immiscible Pb-Sn alloys. As shown in Fig.~\ref{fig_PbSn} (orange circles), the value of $E_j({\rm PbSn})$ is insensitive to the structures and is slightly higher than zero. We thus expect that they can form alloys when they are placed on appropriate substrates: the energy gain due to an electrostatic interaction with substrates overcomes the energy loss due to an alloying of immiscible metals. 

To study the effect of Ag and Al substrates on the Pb-Sn alloying, we show $E_j({\rm PbAg})$, $E_j({\rm PbAl})$, $E_j({\rm SnAg})$, and $E_j({\rm SnAl})$ in Fig.~\ref{fig_Ag_Al}. The B2 shows the lowest value of $E_j$ (except SnAl), while $E_j$s are positive. In particular, PbAl has relatively high value of $E_j\simeq 0.45$ eV. These indicate that Pb and Sn are also immiscible with Al. Therefore, the presence of Al surface will not allow the alloying of Pb and Sn metals, which is consistent with recent observations \cite{yuhara_Al}. Compared with Al alloys, the Ag alloys have lower values of $E_j$. This may yield disordered phase of Pb-Sn on Ag surfaces \cite{yuhara_Ag}. 

Let us move on to the Rh and Ru substrates. Figure \ref{fig_Rh_Ru} shows $E_j({\rm PbRh})$, $E_j({\rm PbRu})$, $E_j({\rm SnRh})$, and $E_j({\rm SnRu})$. For all alloys, the B2 shows the lowest $E_j$. Except PbRu, the value of $E_{\rm B2}(XY)$ is negative, which leads to alloying of elements $X$ and $Y$ with a CN of eight. This would allow the alloying of Pb-Sn on both Rh and Ru surfaces, where the CN is reduced compared with that in bulk. These are consistent with experimental syntheses of Pb-Sn alloys on these surfaces \cite{yuhara_Rh,yuhara_Ru}. The present study shows that the values of $E_{\rm B2}$ of SnRh and SnRu are negatively large compared to those of PbRh and PbRu. This indicates that the creation of Pb-Sn alloys on Rh and Ru surfaces \cite{yuhara_Rh,yuhara_Ru} is mainly due to the strong bonding between Sn atom and Rh or Ru atom in the substrate. 

We try to distinguish ordered and disordered Pb-Sn created on different surfaces. This can be understood by comparing the structures of three-dimensional alloys that have already been synthesized. The information of materials synthesis is extracted from Materials Project database \cite{materialsproject}. For Sn-Ag alloys, SnAg$_3$, i.e., Ag rich alloy, has been synthesized only. On the other hand, for Pb-Rh, Sn-Rh, and Sn-Ru alloys, Pb and Sn rich structures have been reported: for example, Pb$_5$Rh$_4$, Pb$_2$Rh, Sn$_4$Rh, Sn$_2$Rh, Sn$_7$Ru$_3$, and Sn$_3$Ru$_2$. This means that for the former system the concentration of Sn atoms on the Ag surface must be small, rendering it difficult to produce ordered phase. 


We predict appropriate substrate for the Pb-Sn alloying by using the energy of alloys obtained from our previous calculations \cite{ono2020}. Figures \ref{fig_X}(a) and \ref{fig_X}(b)  show the atomic number dependence of $E_j ({\rm Pb}X)$ and $E_j ({\rm Sn}X)$, respectively, for the B2 and B$_h$ structures, where green dashed lines indicate the atomic number of alkali metals. The Li-based alloys have negatively large $E_j$, which may be due to the lightest metallic element. When mixed with heavier alkali metals (K, Rb, and Cs), the values of $E_j$ are nearly zero or positive. For elements heavier than K, $E_j$ behaves periodically: the Pb-based and Sn-based alloys with the group 2 or 3 elements have the minimum $E_j$, while those with the group 6 elements have the maximum $E_j$, followed by a decrease in $E_j$ with increasing the atomic number of $X$. While the difference of $E_j$ between $j=$ B2 and B$_h$ is small compared to the absolute value of $E_j$, the B2 phase seems to be preferable to the B$_h$ phase when $E_j<0$. Table \ref{table1} lists the alloys of Pb$X$ and Sn$X$ that have negative $E_{\rm B2}$. Note that PbCo (0.87), PbIr (0.63), PbNi (0.53), and PbTi (0.31) have positively large value of $E_{\rm B2}$ that overcomes negative $E_{\rm B2}$ of Sn-based alloys. We thus conclude that Au, Ba, Ca, Hf, Li, Lu, Mg, Na, Pd, Pt, Rh, Ru, Sc, Sr, Y, and Zr are suitable substrates for synthesizing Pb-Sn alloys. It is noteworthy that the growth of Sn on the Pt(111) has already been reported \cite{overbury}, so that it would be interesting to study how the addition of Pb atoms influences the structure of the Sn-Pt surface alloy. We also note that some impurities will be needed to stabilize the surface of alkali metals (Li and Na). 



\begin{table}[bb]
\begin{center}
\caption{Formation energy (in units of eV) of Pb-based and Sn-based alloys in the B2 structure.}
{
\begin{tabular}{cccccccccc}\hline
     & Au     &  Ba    &  Ca   & Co       & Hf       & Ir       & Li      & Lu       \\
     \hline
Pb & -0.12 & -1.03  & -1.10 & +0.87  & +0.27 & +0.63 & -0.63 & -0.74  \\
Sn & -0.19 & -1.13  & -1.25  & -0.09  & -0.38  & -0.41  & -0.72 & -1.16   \\
\hline
     &  & & &  & & &  & & \\ 
\hline
     & Mg  & Na     &  Ni    &  Pd   & Pt       & Rh       & Ru       & Sc       \\
     \hline
Pb & -0.17 & -0.31 & +0.53  & -0.40 & -0.09  & -0.11 & +0.39 & -0.65  \\
Sn & -0.29 & -0.29 & -0.13  & -0.84  & -0.61  & -1.00  & -0.76  & -1.07   \\
\hline
     &  & & &  & & &  & & \\ 
\hline
     & Sr     & Ti     & Y     &  Zr     \\
     \hline
Pb & -1.07 & +0.31 & -0.98 & -0.08   \\
Sn & -1.20 & -0.31 & -1.35 & -0.66    \\
\hline
\end{tabular}
}
\label{table1}
\end{center}
\end{table}

It must be noted that our approach relies on negative $E_j$ between the surface metals and the substrate: in the present case, $E_j({\rm SnRh})$ and $E_j({\rm SnRu})$ are negative. The present approach can be applied to understand LiMg alloying on Cu(001) \cite{chen}. While Li-Mg alloys have not been synthesized in the bulk forms, negative values of $E_{\rm B2}({\rm LiCu})$ (-0.05 eV) and $E_{\rm B2}({\rm MgCu})$ (-0.27 eV), where the values of $E_j$ are referred to Ref.~\cite{ono2020}, explain the LiMg alloying on the Cu surface. 

\subsection{Anomalous surface alloys}
\label{subsec:anomalous}
\begin{figure}[tt]
\center
\includegraphics[scale=0.5]{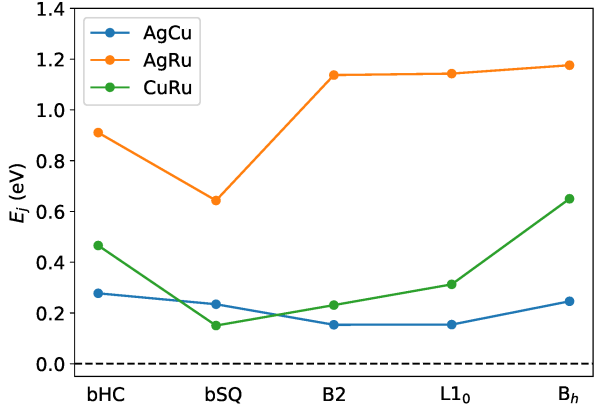}
\caption{Same as Fig.~\ref{fig_PbSn} but for AgCu, AgRu, and CuRu.  } \label{fig_Ag_Cu} 
\end{figure}

\begin{figure}[tt]
\center
\includegraphics[scale=0.5]{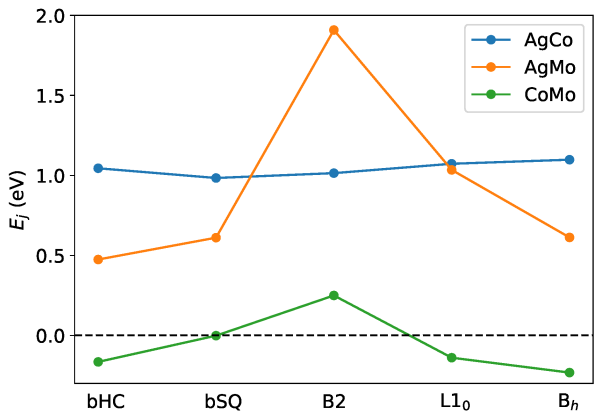}
\caption{Same as Fig.~\ref{fig_PbSn} but for AgCo, AgMo, and CoMo.  } \label{fig_Ag_Co} 
\end{figure}

Within the present approach it is difficult to understand the surface alloying when all three elements are immiscible \cite{steve,sadigh}. For example, we consider AgCu on Ru(0001), where Ag, Cu, and Ru are immiscible with each other \cite{steve}. In this alloy, the lattice constant of Ag and Cu in the alloy structure is similar to that of Ru substrate. The strain energy is thus reduced significantly, providing the energetic stability of this surface alloy \cite{steve}. For an alloy driven by the strain relief, the structure-dependence of $E_j$ is different from that in Pb-Sn alloys, as shown in Fig~\ref{fig_Ag_Cu}. The value of $E_{\rm bSQ}$ is the lowest in Ag-Ru and Cu-Ru alloys, implying that the small CN condition (or small interatomic distance) is preferred. 



Different profile of $E_j$ is also obtained in the immiscible Ag-Co alloys on the Mo(110) surface \cite{tober}, where Ag is immiscible with Co and Mo. Tober et al. have observed the stripe structure of Ag and Co on the Mo(110) with a period of a few nanometers and interpreted that the presence of the stripe is preferable to reduce the misfit strain (or the net area of the surface that consists of Ag and Co) \cite{tober}. As shown in Fig.~ \ref{fig_Ag_Co}, $E_{\rm B2}({\rm AgMo})$ and $E_{\rm B2}({\rm CoMo})$ have the maximum value among five structures. Therefore, the surface alloying observed in the experiment may be related to the stability of bHC or B$_h$ phases because $E_{\rm bHC}({\rm CoMo})$ and $E_{{\rm B}_h}({\rm CoMo})$ are negative. 

In Ref.~\cite{tersoff}, Tersoff has demonstrated that the clusters or the stripe structure at the surface will be created by the presence of a large interface energy with the use of a simplified model taking into account the strain energy only. This is consistent with the observations in AgCu alloy on Ru(0001) for large concentration of Cu \cite{steve}, where pure Ag domain walls are present, and the observed stripe structure in AgCo alloy on Mo(110) \cite{tober}. In order to treat strain effects effectively, metastability analysis used in the present work should be further extended by considering other structures with a relatively large unit cell, which will provide a comprehensive understanding of immiscibility of metals. 

\section{Conclusion}
We have proposed the modified formation energy, Eq.~(\ref{eq:Eform2}), to understand the effect of different CNs on the alloying. By analyzing the formation energies of binary alloys in various structures with different CNs, we have explained the alloying properties of immiscible metals (Pb and Sn) on another metal surface (Rh, Ru, Ag, and Al) \cite{yuhara_Rh,yuhara_Ru,yuhara_Ag,yuhara_Al}. The negatively large formation energy of the B2 structure (CN$=$8) would be important for predicting surface alloying of immiscible metals. The syntheses of Pb-Sn alloys at the predicted substrates are left for future work. We have also identified anomalous surface alloying in our metastability perspective. 

In a future work, we will search appropriate descriptors that predict surface alloying accurately, beyond Eq.~(\ref{eq:Eform2}). It would be fundamentally important to understand the relative stability between metastable structures (i.e., B2, L1$_0$, B$_h$, and other complex structures) under different conditions (i.e., pressure and temperature). For the ground state structure of alloys, it has been known that the combination of pseudo-potential orbital radii is a good descriptor to distinguish complex structures but is not suitable to distinguish B2 and L1$_0$ structures \cite{zunger}. For binary compounds in the zincblend and rocksalt structures, many descriptors have been proposed recently with the help of materials informatics methods \cite{ghi,pilania}. We expect that more research along these lines allows us to identify surface alloying in detail. 

\begin{acknowledgments}
The computation was carried out using the facilities of the Supercomputer Center, the Institute for Solid State Physics, the University of Tokyo, and using the supercomputer ``Flow'' at Information Technologcy Center, Nagoya University.
\end{acknowledgments}






\begin{thebibliography}{99}

\bibitem{pettifor} D. Pettifor, {\it Bonding and Structure of Molecules and Solids} (Oxford University Press, New York, 2002).

\bibitem{ceder} S. Curtarolo, D. Morgan, K. Persson, J. Rodgers, and G. Ceder, Predicting Crystal Structures with Data Mining of Quantum Calculations, Phys. Rev. Lett. {\bf 91}, 135503 (2003).

\bibitem{schleder} G. R. Schleder, A. C. M. Padilha, C. M. Acosta, M. Costa, and A. Fazzio, From DFT to machine learning: recent approaches to materials science-a review, J. Phys.: Mater. {\bf 2}, 032001 (2019).

\bibitem{wolverton} C. Walverton and V. Ozoli\ifmmode \mbox{\c{n}}\else \c{n}\fi{}\ifmmode \check{s}\else \v{s}\fi{}, First-principles aluminum database: Energetics of binary Al alloys and compounds, Phys. Rev. B {\bf 73}, 144104 (2006). 

\bibitem{hart} G. L. W. Hart, S. Curtarolo, T. B. Massalski, and O. Levy, Comprehensive Search for New Phases and Compounds in Binary Alloy Systems Based on Platinum-Group Metals, Using a Computational First-Principles Approach, Phys. Rev. X {\bf 3}, 041035 (2013).

\bibitem{nelson_CE} L. J. Nelson and G. L. W. Hart, and S. Curtarolo, Ground-state characterizations of systems predicted to exhibit $L1_1$ or $L1_3$ crystal structures, Phys. Rev. B {\bf 85}, 054203 (2012).

\bibitem{seko} A. Seko, K. Shitara, and I. Tanaka, Efficient determination of alloy ground-state structures, Phys. Rev. B {\bf 90}, 174104 (2014).

\bibitem{overbury} S. H. Overbury and Yi-sha Ku, Formation of stable, two-dimensional alloy-surface phases: Sn on Cu(111), Ni(111), and Pt(111), Phys. Rev. B {\bf 46}, 7868 (1992).

\bibitem{dhaka} R. S. Dhaka, A. K. Shukla, K. Horn, and S. R. Barman, Photoemission study of Al adlayers on Mn, Phys. Rev. B {\bf 84}, 245404 (2011). 

\bibitem{sad} P. Sadhukhan, S. Barman, T. Roy, V. K. Singh, S. Sarkar, A. Chakrabarti, and S. R. Barman, Electronic structure of Au-Sn compounds grown on Au(111), Phys. Rev. B {\bf 100}, 235404 (2019). 


\bibitem{nielsen} L. P. Nielsen, F. Besenbacher, I. Stensgaard, E. Laegsgaard, C. Engdahl, P. Stoltze, K. W. Jacobsen, and J. K. N\o{}rskov, Initial growth of Au on Ni(110): Surface alloying of immiscible metals, Phys. Rev. Lett. {\bf 71}, 754 (1993).

\bibitem{roder} H. R\"oder, R. Schuster, H. Brune, and K. Kern, Monolayer-confined mixing at the Ag-Pt(111) interface, Phys. Rev. Lett. {\bf 71}, 2086 (1993).

\bibitem{nagl} C. Nagl, M. Pinczolits, M. Schmid, P. Varga, and I. K. Robinson, $p$($n$\ifmmode\times\else\texttimes\fi{}1) superstructures of Pb on Cu(110), Phys. Rev. B {\bf 52}, 16796 (1995). 

\bibitem{steve} J. L. Stevens and R. H. Hwang, Strain Stabilized Alloying of Immiscible Metals in Thin Films, Phys. Rev. Lett. {\bf 74}, 2078 (1995). 

\bibitem{tober} E. D. Tober, R. F. C. Farrow, R. F. Marks, G. Witte, K. Kalki, and D. D. Chambliss, Self-Assembled Lateral Multilayers from Thin Film Alloys of Immiscible Metals, Phys. Rev. Lett. {\bf 81}, 1897 (1998). 

\bibitem{sadigh} B. Sadigh, M. Asta, V. Ozoli\ifmmode \mbox{\c{n}}\else \c{n}\fi{}\ifmmode \check{s}\else \v{s}\fi{}, A. K. Schmid, N. C. Bartelt, A. A. Quong, and R. Q. Hwang, Short-Range Order and Phase Stability of Surface Alloys: PdAu on Ru(0001), Phys. Rev. Lett. {\bf 83}, 1379 (1999). 

\bibitem{chen} M.-S. Chen, S. Mizuno, and H. Tochihara, Ordered mixed surface structures formed on Cu(001) by coadsorption of dissimilar metals: (2$\sqrt{2}\times\sqrt{2}$)R45$^\circ$ by Mg and Li, and ($\sqrt{5}\times\sqrt{5}$)R26.7$^\circ$ by Mg and K(Cs), Surf. Sci. {\bf 486}, L480 (2001).

\bibitem{yuhara_Rh} J. Yuhara, M. Schmid, and P. Varga, Two-dimensional alloy of immiscible metals: Single and binary monolayer films of Pb and Sn on Rh(111), Phys. Rev. B {\bf 67}, 195407 (2003). 

\bibitem{yuhara_Ru} J. Yuhara, Y. Ishikawa, and T. Matsui, Two-dimensional alloy of immiscible Pb and Sn atoms on Ru(0001), Surf. Sci. {\bf 616}, 131 (2013).

\bibitem{yuhara_Ag} J. Yuhara and T. Ako, Two-dimensional Pb-Sn alloy monolayer films on Ag(111), Appl. Surf. Sci. {\bf 351}, 83 (2015).

\bibitem{yuhara_Al} J. Yuhara and Y. Shichida, Epitaxial growth of two-dimensional Pb and Sn films on Al(111), Thin Solid Films {\bf 616}, 618 (2016).

\bibitem{yang} B. Yang, T. Muppidi, V. V. Ozoli\ifmmode \mbox{\c{n}}\else \c{n}\fi{}\ifmmode \check{s}\else \v{s}\fi{}, and M. Asta, First-principles theory of nanoscale pattern formation in ultrathin alloy films: A comparative study of Fe-Ag on Ru(0001) and Mo(110) substrates, Phys. Rev. B {\bf 77}, 205408 (2008).

\bibitem{marathe2009} M. Marathe, M. Imam, and S. Narasimhan, Elastic and chemical contributions to the stability of magnetic surface alloys on Ru(0001), Phys. Rev. B {\bf 79}, 085413 (2009).

\bibitem{marathe2013} M. Marathe, A. D\'{\i}az-Ortiz, and S. Narasimhan, Ab initio and cluster expansion study of surface alloys of Fe and Au on Ru(0001) and Mo(110): Importance of magnetism, Phys. Rev. B {\bf 88}, 245442 (2013). 

\bibitem{ono2020} S. Ono and H. Satomi, arXiv:2012.04790. 













\bibitem{qe} P. Giannozzi {\it et al.}, Advanced capabilities for materials modelling with Quantum ESPRESSO, J. Phys.: Condens. Matter {\bf 29}, 465901 (2017).

\bibitem{pbe} J. P. Perdew, K. Burke, and M. Ernzerhof, Generalized Gradient Approximation Made Simple, Phys. Rev. Lett. {\bf 77}, 3865 (1996).

\bibitem{dalcorso} A. Dal Corso, Pseudopotentials periodic table: From H to Pu, Computational Material Science {\bf 95}, 337 (2014).

\bibitem{MK} H. J. Monkhorst and J. D. Pack, Special points for Brillouin-zone integrations, Phys. Rev. B {\bf 13}, 5188 (1976).

\bibitem{smearing} N. Marzari, D. Vanderbilt, A. De Vita, and M. C. Payne, Thermal Contraction and Disordering of the Al(110) Surface, Phys. Rev. Lett. {\bf 82}, 3296 (1999).
 


\bibitem{materialsproject} A. Jain, S. P. Ong, G. Hautier, W. Chen, W. D. Richards, S. Dacek, S. Cholia, D. Gunter, D. Skinner, G. Ceder, K. A. Persson, The Materials Project: A materials genome approach to accelerating materials innovation, APL Materials, {\bf 1}, 011002 (2013).

\bibitem{tersoff} J. Tersoff, Surface-Confined Alloy Formation in Immiscible Systems, Phys. Rev. Lett. {\bf 74}, 434 (1995).

\bibitem{zunger} A. Zunger, Structural Stability of 495 Binary Compounds, Phys. Rev. Lett. {\bf 44}, 582 (1980).

\bibitem{ghi} L. M. Ghiringhelli, J. Vybiral, S. V. Levchenko, C. Draxl, and M. Scheffler, Big Data of Materials Science: Critical Role of the Descriptor, Phys. Rev. Lett. {\bf 114}, 105503 (2015).

\bibitem{pilania} G. Pilania, J. E. Gubernatis, and T. Lookman, Classification of octet AB-type binary compounds using dynamical charges: A materials informatics perspective, Sci. Rep. {\bf 5}, 17504 (2015). 


\end{thebibliography}
\end{document}